\documentclass[12pt]{iopart}

\usepackage{cite}
\usepackage{iopams}
\usepackage{graphics} 
\usepackage{color}
\usepackage{epsfig}
\usepackage{subfigure}

\begin{document}

\title[Valence-bond crystalline order: $s=1/2$ $J_{1}$--$J_{2}$ model on the honeycomb lattice]
{Valence-bond crystalline order in the $s=1/2$ $J_{1}$--$J_{2}$ model on the honeycomb lattice}

\author{R F Bishop$^{1}$, P H Y Li$^{1}$ and C E Campbell$^{2}$}
\address{$^{1}$ School of Physics and Astronomy, Schuster Building, The University of Manchester, Manchester, M13 9PL, UK}
\address{$^{2}$ School of Physics and Astronomy, University of Minnesota, 116 Church Street SE, Minneapolis, Minnesota 55455, USA}

\begin{abstract}
  Using the coupled cluster method we study the phase diagram of the
  spin-$1/2$ Heisenberg antiferromagnet on a honeycomb lattice with
  nearest-neighbour exchange coupling $J_{1}>0$ and frustrating
  next-nearest-neighbour coupling $J_{2} \equiv xJ_{1}>0$.  In the
  range $0<x<1$ we find four phases exhibiting respectively N\'{e}el,
  6-spin plaquette, staggered dimer, and N\'{e}el-II orderings, with quantum critical points at $x_{c_{1}} \approx
  0.207(3),$ $x_{c_{2}} \approx 0.385(10)$, and $x_{c_{3}} \approx
  0.65(5)$.  The transitions at $x_{c_{1}}$ and $x_{c_{3}}$ appear to
  be continuous (and hence deconfined) ones, while that at $x_{c_{2}}$
  appears to be a direct first-order one.
\end{abstract}


\section{Introduction}
\label{intro}
Frustrated quantum spin models on the two-dimensional (2D) honeycomb
lattice have become the objects of intense study.  Quantum
fluctuations on spin lattices are generally larger for lower
dimensionality $D$ and smaller values of the coordination number $z$
of the lattice, as well as for smaller values of the spin quantum
number $s$ of the lattice spins.  Spin-1/2 models on the honeycomb
lattice (with $D=2$ and $z=3$) are thus expected to have large quantum
fluctuations, which, in turn, open up the theoretical possibility of
realizing exotic ground-state (GS) phases with novel magnetic
properties and/or novel ordering.

Additional impetus for studying 2D honeycomb models came from the
reported presence of a quantum spin-liquid (QSL) phase in both the
exactly soluble (albeit somewhat artificial) Kitaev model of spin-1/2
particles on a honeycomb lattice \cite{kitaev}, and the half-filled
Fermi-Hubbard (FH) model on a honeycomb lattice \cite{meng}.  Thus,
Meng {\it et al.} \cite{meng} reported in a quantum Monte Carlo (QMC) calculation,
free of the usual fermion sign problems, the presence in the honeycomb
FH model of a QSL phase, at moderate values of the on-site Coulomb
repulsion strength ($U$), situated between the nonmagnetic metallic
insulator (or semi-metal) phase at low $U$ and the antiferromagnetic
(AFM) Mott insulator phase for large $U$.  Since the $U \rightarrow
\infty$ limit corresponds to the pure Heisenberg antiferromagnet
(HAFM), i.e., with nearest-neighbour (NN) interactions (of strength
$J_{1} > 0$) only, the Mott insulator phase of the Hubbard model
corresponds to the N\'{e}el-ordered phase of the HAFM spin-lattice
model.  Higher-order terms in the $t/U$ expansion of the FH model
(where {\it t} is the strength parameter of the NN hopping term) lead
to frustrating exchange couplings in the corresponding spin-lattice
model in which the HAFM with NN exchange couplings is the leading term
in the large-$U$ expansion.  The simplest such frustrated model is the
$J_{1}$--$J_{2}$ model studied here, where the next-nearest-neighbour
(NNN) spin pairs have an additional exchange coupling of strength
$J_{2}>0$.

A later study of the FH model, using a Schwinger boson mean field
theory (SB-MFT) approach \cite{Wang:2010_honey}, provided some
corroborating evidence for a $\mathbb{Z}_{2}$ QSL state; and a
Schwinger fermion representation of the same model
\cite{Lu:2011_honey} gave some evidence for both a $\mathbb{Z}_{2}$
QSL phase and a chiral antiferromagnetic phase.  However, later
numerically exact QMC calculations by Sorella {\it et al.}
\cite{Sorella:2012}, with much larger clusters than those used by Meng
{\it et al.} \cite{meng}, have cast considerable doubt on their
original finding of an intermediate QSL phase.  We note in this
context that the presence of magnetically ordered phases is difficult
to detect by standard QMC techniques when the ordering is small, since
the usual quantity measured is the {\it square} of the order
parameter.  As a consequence, in addition to the usual problem of
finding an appropriate finite-size extrapolation formula, very large
clusters are required with high precision.  It is this effect that has
apparently caused the controversy between Refs.\ \cite{meng} and
\cite{Sorella:2012} regarding the existence or not of an intermediate
QSL phase in the FH model on the honeycomb lattice.  In a very recent
paper \cite{Assaad:2013_honey_Hubbard} this controversy has
effectively been resolved by using a novel QMC technique that measures
the local magnetic order parameter $M$ directly, rather than its
square, $M^{2}$.  Use of this technique leads
\cite{Assaad:2013_honey_Hubbard} to the rather firm conclusion that in
the FH model on the honeycomb lattice there is a single continuous
quantum phase transition between the nonmagnetic semi-metal and AFM
Mott insulator phases, with no intermediate QSL phase.

It is also pertinent to ask whether the $J_{1}$--$J_{2}$ model
actually does represent well the low-energy physics of the FH model on
the honeycomb lattice.  While this is undoubtedly true for small
enough values of the Hubbard parameter $t/U$, it is interesting to
enquire more deeply and quantitatively about this question.  In
particular, two recent studies \cite{Yang:2011,Yang:2012} have thrown
considerable light on the relationship between the physics of FH and
$J_{1}$--$J_{2}$ models on the honeycomb lattice.  Thus, in the first
place, it has been shown \cite{Yang:2011} that the ratio $x \equiv
J_{2}/J_{1}$ actually stays quite small over a large range of values
of $t/U$.  More specifically, it is always smaller than the value
$x_{c_{1}}$, which is the point at which the N\'{e}el order, present
at $x=0$, first vanishes as $x$ is increased, as we discuss below.
Secondly, in a very interesting paper \cite{Yang:2012} that studied in
detail the full low-energy spin model arising from the FH model on the
honeycomb lattice, it was shown that six-spin interactions on
hexagonal plaquettes are the most important leading correction to the
NN $J_{1}$ bonds, rather than the NNN $J_{2}$ bonds.

Despite all of the above caveats of the relevance of the
$J_{1}$--$J_{2}$ model on the honeycomb lattice to describe the
low-energy physics of the corresponding FH honeycomb model, it remains
of very great interest in its own right.  This has possibly even been
heightened by the considerable uncertainty that has existed until very recently, as discussed above, as to
whether or not a QSL phase exists for the FH model.  For this and
other reasons, this spin-lattice model and its generalizations [specifically to
include also next-next-nearest-neighbour (NNNN) bonds with strength
$J_{3}$], have been much studied
\cite{Mulder:2010_honey,DJJF:2011_honeycomb,Albuquerque:2011_honey,Oitmaa:2011_honey,Reuther:2011_honey,Mezzacapo:2012_honey,Bishop:2012_honeyJ1-J2,Li:2012_honey_full,Zhu:2012_honeyJ1-J2,Ganesh:2013_J1J2honey,Zhang:2013_J1J2honey,Yu:2013:honey}
recently.

\section{The model}
\label{model_section}
The Hamiltonian of the model studied here is given by
\begin{equation}
H = J_{1}\sum_{\langle i,j \rangle} \mathbf{s}_{i}\cdot\mathbf{s}_{j} + J_{2}\sum_{\langle\langle i,k \rangle\rangle} 
\mathbf{s}_{i}\cdot\mathbf{s}_{k}\,,
\label{eq1}
\end{equation}
where index $i$ runs over all honeycomb lattice sites, and indices $j$ and $k$ run over all
NN and NNN sites to $i$, respectively, counting each bond once only.  Each lattice site $i$ carries a particle with spin 
$s=\frac{1}{2}$ and a spin operator ${\bf s}_{i}=(s_{i}^{x},s_{i}^{y},s_{i}^{z})$.

\begin{figure}[!tb]
\begin{center}
\mbox{
\subfigure[]{\scalebox{0.35}{\includegraphics{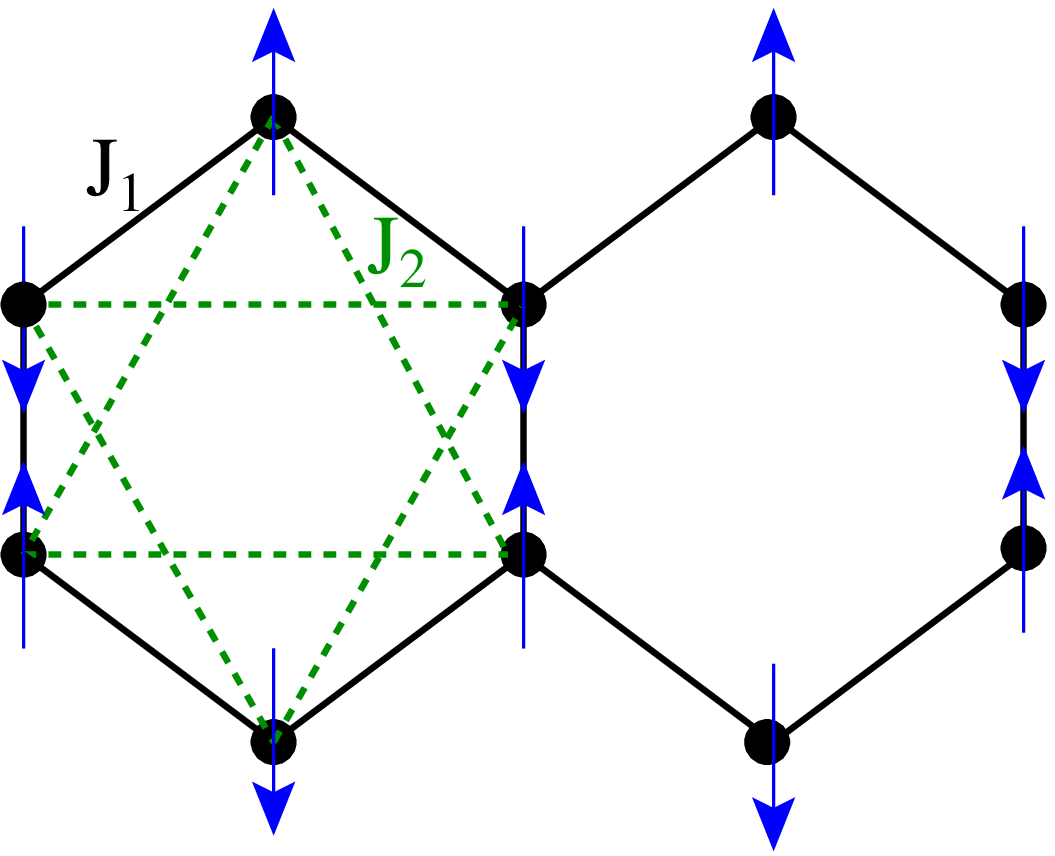}}}
\quad
\subfigure[]{\scalebox{0.35}{\includegraphics{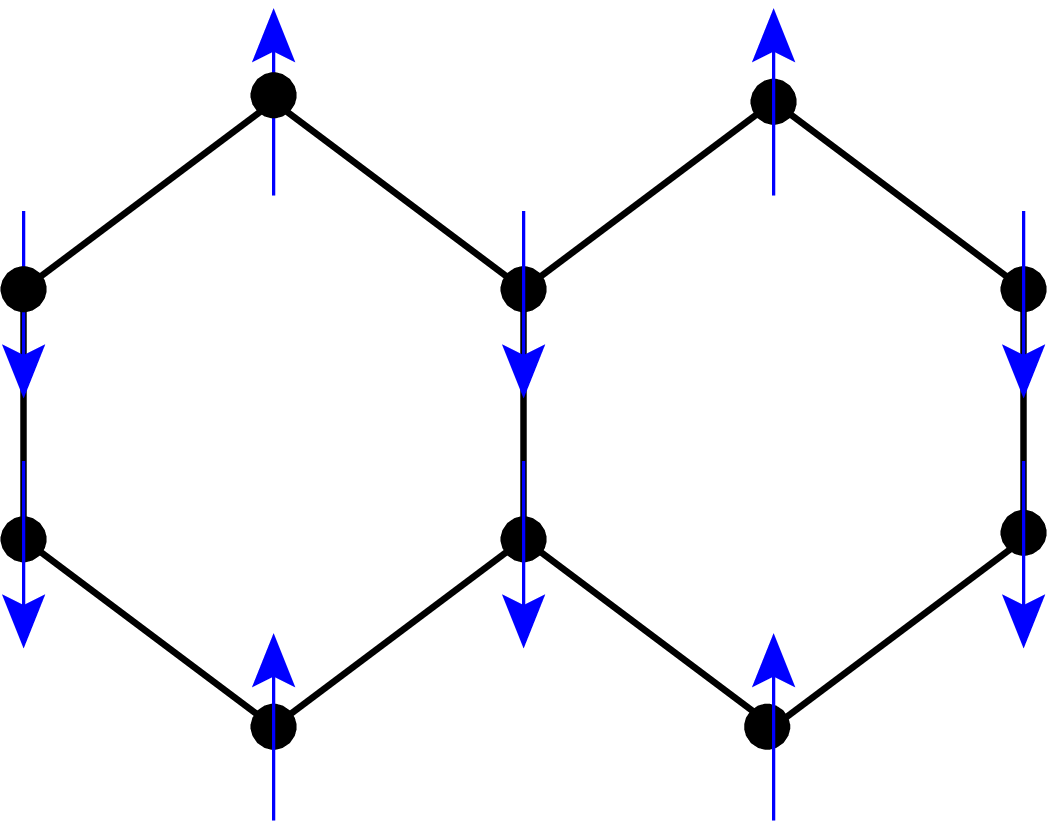}}}
}
\caption{(Colour online) The $J_{1}$--$J_{2}$ model on the honeycomb lattice (with $J_{1}=1$),
  showing (a) the N\'{e}el and (b) N\'{e}el-II states. 
  The arrows represent spins located on lattice sites \textbullet.}
\label{model}
\end{center}
\end{figure}

The lattice and exchange bonds are illustrated in figure~\ref{model}.
We are interested in the case where both NN and NNN bonds are AFM
in nature, and henceforth, we put $J_{1}=1$ to set the energy scale and define
the frustration parameter $x \equiv J_{2}/J_{1}$.

The classical ($s \rightarrow \infty$) ground state of the model is
N\'{e}el-ordered for $0 \leq x < \frac{1}{6}$, whereas for all values
$x > \frac{1}{6}$ the spins are spirally ordered.  In this latter
regime, the classical model has a one-parameter family of degenerate
incommensurate ground states where the spiral wave vector can orient
in any direction.  At leading order, i.e., $O(1/s)$, spin-wave
fluctuations lift this accidental degeneracy in favour of particular
wave vectors \cite{Mulder:2010_honey}.  For the extreme quantum case,
$s=1/2$, considered here, we expect quantum fluctuations to be strong
enough to destroy the spiral order over a wide range of values of $x$.
In a recent paper \cite{Bishop:2012_honeyJ1-J2} that used the coupled
cluster method (CCM), we have verified that expectation for all values
in the range $0 \leq x \leq 1$ considered here.

We showed too \cite{Bishop:2012_honeyJ1-J2} that
quantum fluctuations preserve the N\'{e}el order to higher values of
$x$ than in the classical model.  Thus, we found that the GS phase
of the $s=1/2$ model is N\'{e}el-ordered for $x<x_{c_{1}} \approx
0.207(3)$.  At $x=x_{c_{1}}$ there appears to be a continuous
deconfined phase transition to a GS paramagnetic phase exhibiting
plaquette valence-bond crystalline (PVBC) order.  Furthermore, we
found the PVBC state to be the stable GS phase in the regime
$x_{c_{1}}<x<x_{c_{2}}$, where $x_{c_{2}} \approx 0.385(10)$.
Our aim now is to investigate further the transition at $x=x_{c_{2}}$
and the nature of the GS phase(s) for $x>x_{c_{2}}$. 

\section{Coupled cluster method}
\label{CCM}
The CCM \cite{Bi:1991,Bi:1998,Fa:2004}, that we will employ here, has been very successfully applied
to many models in quantum magnetism, including models on the honeycomb
lattice
\cite{DJJF:2011_honeycomb,Bishop:2012_honeyJ1-J2,Li:2012_honey_full}
of interest here.  It provides a well-structured means of studying
various candidate GS phases and their regimes of stability, for each
of which the description is systematically improvable in terms of
well-defined truncation hierarchies for the quantum multi-spin
correlations.  We now briefly describe the method and refer the reader
to the literature (see,
e.g.,~\cite{Bi:1991,Bi:1998,Fa:2004,Ze:1998,Kr:2000,Bi:2000,Darradi:2005,Bi:2008_PRB_J1xxz_J2xxz,Darradi:2008_J1J2mod,Bishop:2010_UJack})
for further details.

The starting point for any CCM calculation is the selection of a
suitable normalized model (or reference) state $|\Phi\rangle$.  For
spin systems it is often convenient to take a classical (uncorrelated)
GS wave function for $|\Phi\rangle$.  For the present case we choose
the N\'{e}el state shown in figure~\ref{model}(a) for small values of
the frustration parameter $x$.  For larger values of $x$ we could
choose one of the classical spiral GS phases to provide a CCM model
state, but as we have argued above these are likely to be very fragile
against quantum fluctuations.  Instead, for larger values of $x$, we
choose here the so-called N\'{e}el-II phase shown in
figure~\ref{model}(b) (which has also been denoted as the
anti-N\'{e}el phase earlier \cite{Bishop:2012_honeyJ1-J2}), that
occurs in the classical ($s\rightarrow\infty$) model only at the
isolated and highly degenerate critical point $x=\frac{1}{2}$.
Whereas the N\'{e}el state has all 3 NN spins to a given spin
antiparallel to it, the N\'{e}el-II state also comprises AFM sawtooth
chains along one of the three equivalent honeycomb directions, but
with NN spins on adjacent chains now parallel to one another.  The
N\'{e}el-II state is also sometimes known in the literature as the
collinear striped AFM phase for reasons that should be clear from
figure~\ref{model}(b), although we prefer to avoid this name here
since it is open to confusion with other AFM states on the honeycomb
lattice that have also been called striped states (see,
e.g.,~\cite{Li:2012_honey_full}).  The N\'{e}el-II state is thus also
easily seen to break the lattice rotational symmetry.

It is convenient to perform a mathematical rotation of the local axes
of the spins such that all spins in the reference state align along
the negative $z$-axis.  The Schr\"{o}dinger ground-state ket and bra
CCM equations are $H|\Psi\rangle = E|\Psi\rangle$ and
$\langle\tilde{\Psi}|H=E\langle\tilde{\Psi}|$ respectively.  The CCM
employs the exponential parametrizations, $|\Psi\rangle={\rm
  e}^{S}|\Phi\rangle$ and
$\langle\tilde{\Psi}|=\langle\Phi|\tilde{S}$e$^{-S}$. The
correlation operator $S$ is expressed as $S = \sum_{I\neq0}{\cal
  S}_{I}C^{+}_{I}$ and its counterpart is $\tilde{S} = 1 +
\sum_{I\neq0}\tilde{\cal S}_{I}C^{-}_{I}$ where, by definition,
$C^{-}_{I}|\Phi\rangle = 0 = \langle\Phi|C^{+}_{I}, \forall I \neq 0$.
Thus we have the normalization condition
$\langle\tilde{\Psi}|\Psi\rangle = \langle\Phi|\Phi\rangle \equiv 1$.
The multispin creation operators $C^{+}_{I} \equiv
(C^{-}_{I})^{\dagger}$, with $C^{+}_{0} \equiv 1$, are written as
\(C^{+}_{I}\equiv s^{+}_{j_{1}} s^{+}_{j_{2}} \cdots s^{+}_{j_{n}}\),
in terms of the single-site spin-raising operators $s^{+}_{k}\equiv
s^{x}_{k}+is^{y}_{k}$.  The GS energy is $E=
\langle\Phi|\mbox{e}^{-S}H\mbox{e}^{S}|\Phi\rangle$; and the local
average onsite magnetization $M$ in the rotated spin coordinates is $M
\equiv -\frac{1}{N}
\langle\tilde{\Psi}|\sum_{j=1}^{N}s^{z}_{j}|\Psi\rangle$.  The ket-
and bra-state correlation coefficients $({\cal S}_{I}, \tilde{{\cal
    S}_{I}})$ are calculated by requiring the expectation value
$\bar{H}=\langle\tilde{\Psi}|H|\Psi\rangle$ to be a minimum with
respect to all parameters $({\cal S}_{I}, \tilde{{\cal S}_{I}})$, and
hence $\langle \Phi|C^{-}_{I}\mbox{e}^{-S}H\mbox{e}^{S}|\Phi\rangle =
0$ and $\langle\Phi|\tilde{S}(\mbox{e}^{-S}H\mbox{e}^{S} -
E_{0})C^{+}_{I}|\Phi\rangle = 0\;; \forall I \neq 0$.

The CCM formalism is exact if all spin configurations are included in
the $S$ and $\tilde{S}$ operators.  In practice, however, truncations
are needed.  We employ here the well-studied localized (lattice-animal-based subsystem) LSUB$m$
scheme~\cite{Ze:1998,Kr:2000,Bi:2000,Darradi:2005,Bi:2008_PRB_J1xxz_J2xxz,Darradi:2008_J1J2mod,Bishop:2010_UJack},
in which all possible multi-spin-flip correlations over different
locales on the lattice defined by $m$ or fewer contiguous lattice
sites are retained.  Such clusters are defined to be contiguous in
this sense if every site in the cluster is adjacent (as a nearest
neighbour) to at least one other site in the cluster.  The interested reader is referred to the literature (see, e.g., \cite{Ze:1998}) for figures illustrating the LSUB$m$ scheme in detail.  The numbers
$N_{f}$ of such fundamental configurations that are distinct under the
(space and point-group) symmetries of the lattice and the model state
increase rapidly with the LSUB$m$
truncation index $m$.  Thus the highest LSUB$m$ level that we
can reach here, even with massive parallelization and the use of
supercomputing resources \cite{ccm}, is LSUB$12$, for which $N_{f} = 293309$ for
the N\'{e}el-II state.  

Since, in any truncation, CCM parametrizations automatically satisfy
the Goldstone linked cluster theorem, we may work from the outset in
the thermodynamic limit, $N \rightarrow \infty$.  Nevertheless, the
raw LSUB$m$ data still need to be extrapolated to the exact $m
\rightarrow \infty$ limit.  Thus, for the GS energy per spin, $E/N$,
we use (see, e.g., \cite{Bi:2000,Darradi:2005,Bi:2008_PRB_J1xxz_J2xxz,Darradi:2008_J1J2mod})
\begin{equation}
E(m)/N = a_{0}+a_{1}m^{-2}+a_{2}m^{-4}\,;     \label{E_extrapo}
\end{equation}
while for the magnetic order parameter, $M$, defined above,
we use either the scheme
\begin{equation}
M(m) = b_{0}+b_{1}m^{-1}+b_{2}m^{-2}\,,    \label{M_extrapo_standard}
\end{equation}
for systems showing no or only slight frustration
(see, e.g., \cite{Kr:2000,Darradi:2005}), or the scheme
\begin{equation}
M(m) = c_{0}+c_{1}m^{-1/2}+b_{2}m^{-3/2}\,,   \label{M_extrapo_frustrated}
\end{equation}
for more strongly frustrated systems or ones showing a GS
order-disorder transition (see, e.g.,
\cite{Bi:2008_PRB_J1xxz_J2xxz,Darradi:2008_J1J2mod}).  

In principle one may always test for the correct leading exponent in
the LSUB$m$ extrapolation scheme for any physical quantity $Z$ by
first fitting to the formula $Z(m)=d_{0}+d_{1}m^{-\nu}$.  For the GS
energy, $E/N$, we generally find $\nu \approx 2$ for a wide
variety of spin systems, both non-frustrated and frustrated.  For the
magnetic order parameter, $M$, on the other hand we generally find $\nu
\approx 1$ for unfrustrated systems (or for ones with very small
frustration), and $\nu \approx 0.5$ for more strongly frustrated
systems.  We discuss this more fully in section \ref{results} in the context of the
present model.  These general results for the leading exponents then provide the basis for equations (\ref{E_extrapo})-(\ref{M_extrapo_frustrated}).

Finally, we note that since the hexagon is an important structural
element of the honeycomb lattice we never use LSUB$m$ data with $m <
6$ to perform the extrapolations.  Furthermore, in any CCM calculation using the LSUB$m$ scheme, we always need to
check whether the lowest-order potentially usable approximation, namely
LSUB6 here, is {\it actually} usable in the sense of fitting the
extrapolation scheme to be used.  Although it generally does do so,
there are also (relatively rare) occasions when it does not,
presumably due either to the result being too far removed from the
asymptotic $m \rightarrow \infty$ limit or to the fact that for the
particular CCM model state used these lowest-order approximants omit
one or more of the most important multispin correlations.

\section{Results and discussion}
\label{results}
\begin{figure}[t]
\begin{center}
\includegraphics[width=6cm,angle=270]{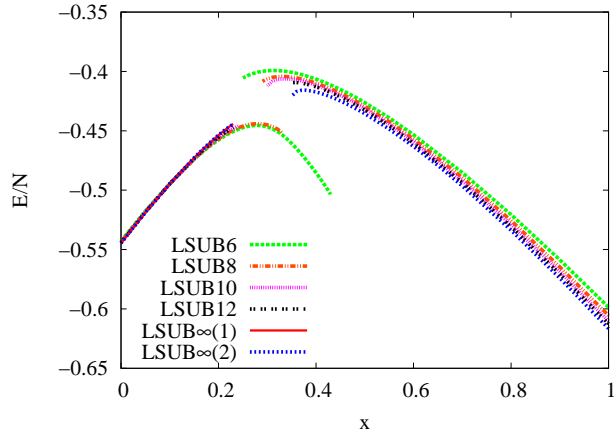}  
\caption{(Colour online) CCM LSUB$m$ results for the GS energy per
  spin, $E/N$, as a function of the frustration parameter, $x \equiv
  J_{2}/J_{1}$, of the spin-$1/2$ $J_{1}$--$J_{2}$ honeycomb model
  (with $J_{1}>0$), using the N\'{e}el (left curves) and N\'{e}el-II
  (right curves) states as model states, with $m=\{6,8,10,12\}$.  The extrapolated curves
  LSUB$\infty$(1) and LSUB$\infty$(2) use this data set and the restricted set
  $m=\{8,10,12\}$ respectively, with
  equation~(\ref{E_extrapo}).}
\label{E}
\end{center}
\end{figure}

\begin{figure}[t]
\begin{center}
\includegraphics[width=6cm,angle=270]{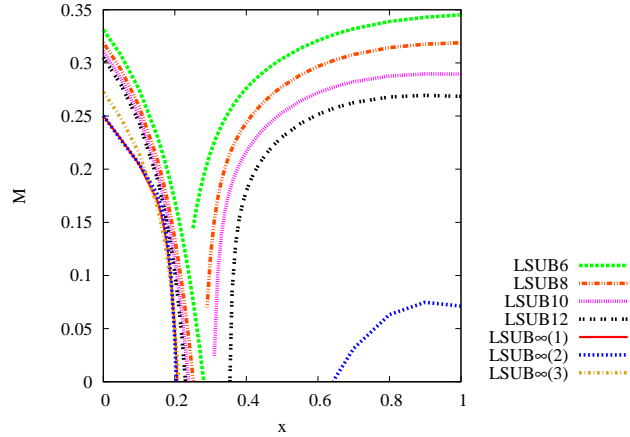}  
\caption{(Colour online) CCM LSUB$m$ results for the GS order parameter, $M$, as a function of the frustration parameter, $x \equiv J_{2}/J_{1}$, of the
  spin-$1/2$ $J_{1}$--$J_{2}$ honeycomb model ($J_{1}=1$), using the
  N\'{e}el (left curves) and N\'{e}el-II (right curves) states as
  model states, with $m=\{6,8,10,12\}$.  The extrapolated curves
  LSUB$\infty$(1) and LSUB$\infty$(3) use this data set with
  equations~(\ref{M_extrapo_frustrated}) and (\ref{M_extrapo_standard})
  respectively, while the LSUB$\infty$(2) curve uses
  equation~(\ref{M_extrapo_frustrated}) with the restricted set
  $m=\{8,10,12\}$.}
\label{M}
\end{center}
\end{figure}

In figures~\ref{E} and \ref{M} we show our results for the GS energy per spin, $E/N$, and magnetic order
parameter, $M$, using both the N\'{e}el and
N\'{e}el-II states as CCM model states.  

Figure~\ref{E} shows clearly that the CCM LSUB$m$ results for the GS
energy extrapolate extremely rapidly with increasing order $m$ of
approximation to the exact LSUB$\infty$ limit.  It also shows clearly
how the LSUB$m$ results based on both the N\'{e}el and N\'{e}el-II model states naturally terminate at some critical values of the frustration
parameter $x$, which themselves depend on the order parameter $m$ of
the particular LSUB$m$ approximation, beyond which no real CCM
solution can be found.  Such termination points of CCM solutions are
well studied (and see, e.g., Refs.\ \cite{Bi:1998,Fa:2004}) and well
understood.  They are simply reflections of the quantum phase
transitions in the system and, as such, may themselves be used to
estimate the positions of the corresponding quantum critical points
\cite{Bi:1998}.  We do not, however, examine the extrapolation
properties of the termination points further here, since we have more
accurate criteria available to us to determine the quantum critical
points, as we discuss more fully below.  Nevertheless, figure \ref{E}
shows clearly that the CCM LSUB$m$ results based on both the N\'{e}el
and N\'{e}el-II model states for finite values of $m$ extend beyond the
corresponding LSUB$\infty$ transition points into unphysical regions
where such states in the real (LSUB$\infty$) case have ceased to
exist.  Such unphysical regimes diminish in size to zero as $m
\rightarrow \infty$.  Figure \ref{E} shows that there are no energy
crossings between the N\'{e}el and N\'{e}el-II phases at any LSUB$m$
level of approximation, and that there is a clear range of values of
the frustration parameter, $x_{c_{1}} < x < x_{c_{2}}$, in which
neither the N\'{e}el nor the N\'{e}el-II states provide a physical GS
phase.  The simple unextrapolated LSUB12 estimates for the two
termination points, namely $x_{c_{1}} \lesssim 0.23$ and $x_{c_{2}}
\gtrsim 0.35$ already provide remarkably good estimates for the
corresponding quantum critical points, as we shall see below.

We note from figure \ref{E} that the LSUB$m$ estimates for the GS energy
approach the asymptotic LSUB$\infty$ limit very rapidly, and hence the
extrapolations are rather insensitive to both the fitting scheme and data set
used.  Nevertheless, a fit of the form $E(m)/N = e_{0} +
e_{1}m^{-\nu}$ for the N\'{e}el-II LSUB$m$ results gives the usual
expected result $\nu \approx 2$ for the data set $m=\{8,10,12\}$,
whereas the inclusion of the LSUB6 result leads to a spurious value
$\nu \approx 1$.  By contrast, both data sets $m=\{6,8,10,12\}$ and
$m=\{8,10,12\}$ yield a value $\nu \approx 2$ for the corresponding
LSUB$m$ N\'{e}el results.  The anomalous nature of the LSUB6
N\'{e}el-II approximation is discussed further below with regard to
the magnetic order parameter $M$, for which its behaviour is more
critical and more pronounced.

We now turn our attention to the corresponding CCM LSUB$m$ results for
the magnetic order parameter, as shown in figure~\ref{M}, using both
the N\'{e}el and N\'{e}el-II states as the CCM model states.  For the
present model we find that an extrapolation formula for the magnetic
order parameter of the form $M(m)=d_{0}+d_{1}m^{-\nu}$ fits the data
well on the N\'{e}el side with a leading exponent $\nu \approx 1$ for
values of the frustration parameter $x$ equal to or very close to
zero, whereas the value $\nu \approx 0.5$ accurately fits the data
over most of the range $x \gtrsim 0.1$.  Accordingly, in
figure~\ref{M} on the N\'{e}el side we show extrapolations using both
equations~(\ref{M_extrapo_standard}) and (\ref{M_extrapo_frustrated}).
Equation (\ref{M_extrapo_standard}),
which is appropriate when $J_{2}=0$, yields the value $M \approx
0.271(2)$ for the unfrustrated HAFM on the hexagonal lattice (i.e.,
with NN interactions only), in excellent agreement with the best
available QMC estimate \cite{Castro:2006_HC}, $M=0.2677(6)$.  Our own
error estimates are based on sensitivity checks using different
LSUB$m$ data sets.  Similarly we see from figure~\ref{M} that all
extrapolations give essentially the same estimate $x_{c_{1}} \approx
0.207(3)$ for the point where N\'{e}el order vanishes ($M \rightarrow
0$).  We showed previously \cite{Bishop:2012_honeyJ1-J2} that the phase
transition at $x=x_{c_{1}}$ is a continuous deconfined one between
states with N\'{e}el and PVBC order.

Figure \ref{M} also shows corresponding results for $M$ for a possible
phase with N\'{e}el-II ordering.  In this case we find (even by simple
inspection by eye) that the LSUB6 results do not fit with a
leading-order extrapolation scheme of the form
$M(m)=d_{0}+d_{1}m^{-\nu}$ with {\it any} value of $\nu$.  By
contrast, the LSUB$m$ results with $m>6$ are accurately fitted by this
form with a leading-order exponent $\nu \approx 0.5$ over the whole
range of values of the frustration parameter $x$ shown.  Precisely why
the LSUB6 result should be anomalous in this case is unclear, but as
discussed in section \ref{CCM} we must now discard it for
extrapolation purposes.  For these reasons we show in figure~\ref{M}
only extrapolated results using equation (\ref{M_extrapo_frustrated})
for the N\'{e}el-II model state, based on $m=\{8,10,12\}$.  The
results clearly show that N\'{e}el-II ordering is present, albeit with
a rather small value of the order parameter, $M \lesssim 0.1$, for
$x>x_{c_{3}}$ where $x_{c_{3}} \approx 0.65(5)$, but where the error
estimate is now more uncertain.

\begin{figure}[t]
\begin{center}
\mbox{
\subfigure{\includegraphics[width=6cm,height=6cm,angle=270]{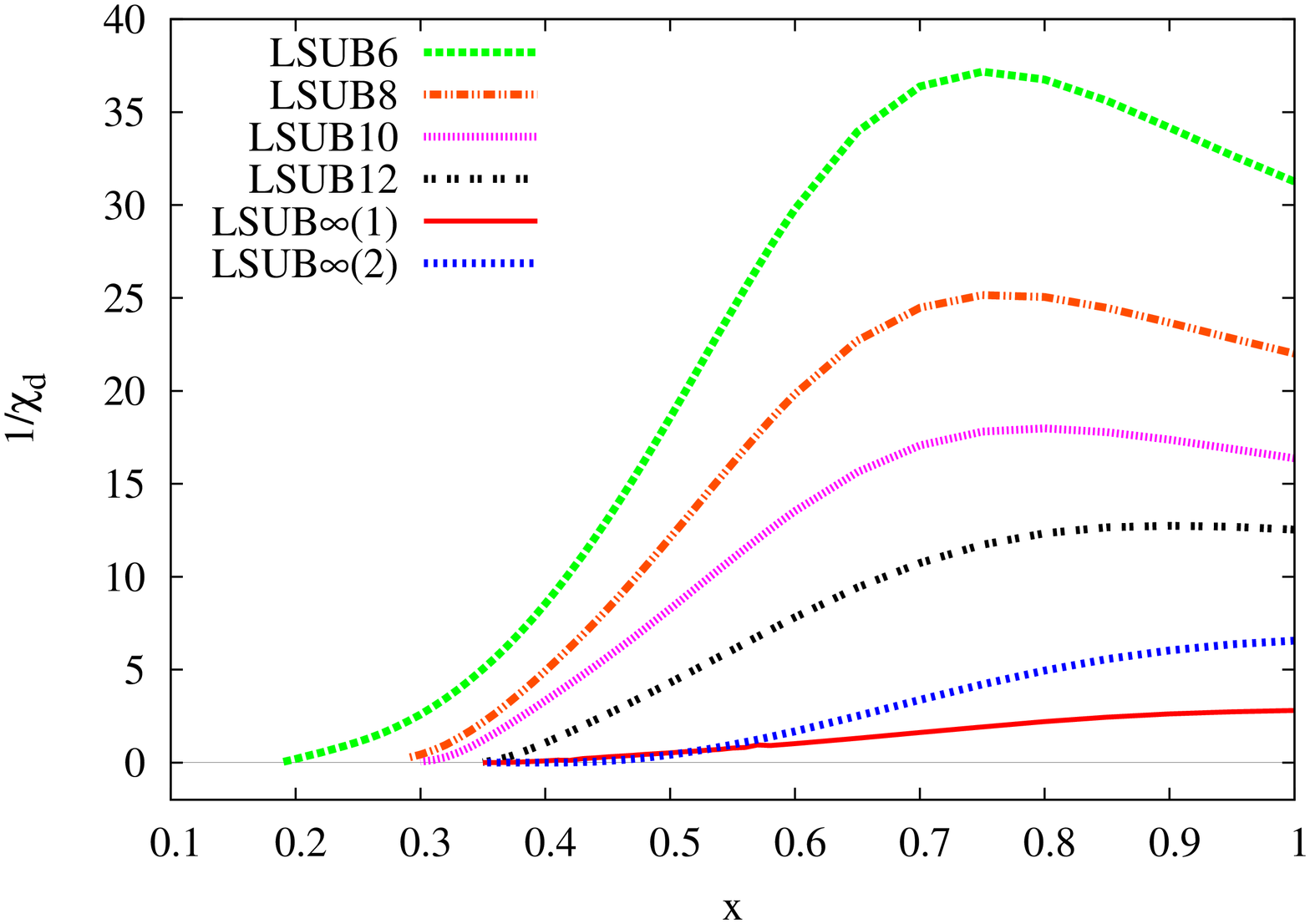}}
\raisebox{-3.5cm}{
\subfigure{\includegraphics[width=2.2cm,height=2.2cm]{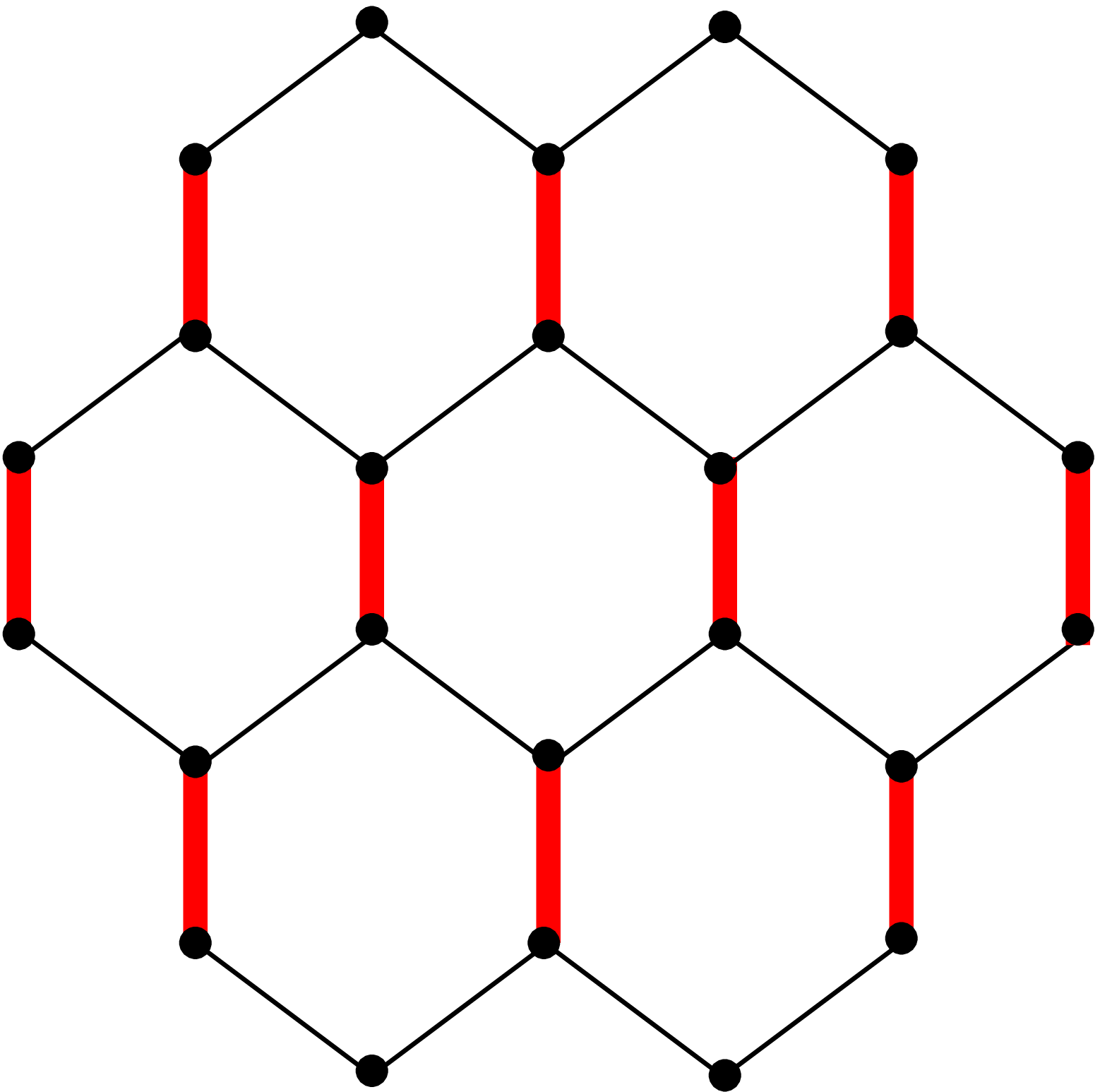}}
}
}
\caption{(Colour online) Left: CCM LSUB$m$ results for the inverse
  staggered dimer susceptibility, $1/\chi_{d}$, as a function of the frustration parameter, $x \equiv J_{2}/J_{1}$, of the
  spin-1/2 $J_{1}$--$J_{2}$ honeycomb model ($J_{1}=1$),
  using the N\'{e}el-II state as model state, with $m=\{6,8,10,12\}$.
  The extrapolated curves LSUB$\infty$(1) and LSUB$\infty$(2) are
  derived from fitting the perturbed energies (see text) as
  $e(\delta)=e_{0}(\delta)+e_{1}(\delta)m^{-\nu}$, and use the data sets
  $m=\{6,8,10,12\}$ and  $m=\{8,10,12\}$ respectively.  Right: The field $F \rightarrow
  \delta\; \hat{O}_{d}$ for the staggered dimer susceptibility,
  $\chi_{d}$.  Thick (red) and thin (black) lines correspond
  respectively to strengthened and unaltered NN exchange couplings,
  where $\hat{O}_{d} = \sum_{\langle i,j \rangle} a_{ij}
  \mathbf{s}_{i}\cdot\mathbf{s}_{j}$, and the sum runs over all NN
  bonds, with $a_{ij}=+1$ and 0 for thick (red) lines and thin
  (black) lines respectively.}
\label{X}
\end{center}
\end{figure}

In our previous work \cite{Bishop:2012_honeyJ1-J2} we showed that the
N\'{e}el-II state becomes susceptible to PVBC ordering for
$x<x_{c_{2}} \approx 0.385(10)$, but we now observe that the
N\'{e}el-II state is itself only stable as a magnetically ordered
state for $x>x_{c_{3}}$.  We are thus led to enquire about the
possible GS phase(s) of the system in the range
$x_{c_{2}}<x<x_{c_{3}}$.  In view of the persistence of our CCM
LSUB$m$ solutions based on the N\'{e}el-II model state, with finite
values of $m$, well into the region $x<x_{c_{3}}$ before they
terminate (as is clearly seen from figure~\ref{M}), we expect that the
actual GS phase in this intermediate regime might share similarities
with the N\'{e}el-II state.  For example, just as the N\'{e}el-II state
breaks the lattice rotational symmetry, so does another valence-bond
solid state, namely the staggered-dimer valence-bond crystalline
(SDVBC) (or lattice nematic) state.  This is formed from the
N\'{e}el-II state by replacing all of the parallel NN spin pairs by spin-zero dimers (and see
figure~\ref{X}).

In order to investigate the possibility of an SDVBC phase we first
consider the
response of the system to a field operator $F$ (and see, e.g., \cite{Darradi:2008_J1J2mod}).  Thus, a field term $F=\delta\;
\hat{O}_{d}$ is added to the Hamiltonian of equation~(\ref{eq1}), where
$\hat{O}_{d}$ is an operator corresponding to the possible SDVBC
order, illustrated in figure~\ref{X} and defined in its caption.  The
energy per site, $E(\delta)/N \equiv e(\delta)$, is then calculated in
the CCM for the perturbed Hamiltonian $H + F$, using the N\'{e}el-II
model state.  We define the corresponding susceptibility as $\chi_{d}
\equiv - \left. (\partial^2{e(\delta)})/(\partial {\delta}^2)
\right|_{\delta=0}$.  Clearly the GS phase becomes unstable against
SDVBC order when $\chi_d^{-1}$ becomes zero.  We now use the LSUB$m$ extrapolation scheme $e(\delta) =
e_{0}(\delta)+e_{1}(\delta)m^{-\nu}$, with the exponent $\nu$ also a
fitting parameter, rather than our standard energy extrapolation
scheme of equation~(\ref{E_extrapo}), in order to calculate the
extrapolated values of $\chi^{-1}_{d}$ shown in figure~\ref{X}.  For the
same data set $m=\{8,10,12\}$ used to calculate $M$ for the
N\'{e}el-II state above, the fitted value of $\nu$ is close to 2 over
most of the range of the $J_{2}$ values shown, except near the
termination point of this phase, where it falls sharply.  By contrast,
for the set $m=\{6,8,10,12\}$ also shown in figure~\ref{X}, $\nu$ is closer to 1 over most of the range.  This again
reinforces the anomalous nature of the LSUB6 results.

What we see from figure~\ref{X} is that the extrapolated value of
$\chi^{-1}_{d}$ is close to zero over a range of values of $x$ that
extends from $x_{c_{2}}$ below to an upper value of about 0.6, which
is completely compatible with the value $x_{c_{3}}$ obtained from the
order parameter $M$ of the N\'{e}el-II state.  Thus, by combining our
results, we conclude that in the region $x_{c_{2}}<x<x_{c_{3}}$ the GS
phase has SDVBC order, while for $x>x_{c_{3}}$ the GS phase has
N\'{e}el-II order, although this latter ordering is weak and quite
fragile against the still strongly competing SDVBC order.  The shape
of the CCM curves for $\chi^{-1}_{d}$ in figure~\ref{X} are indicative
of a continuous (and hence deconfined) quantum critical point at
$x_{c_{3}}$, whereas the corresponding curves for $\chi^{-1}_{p}$, the
inverse plaquette susceptibility, found in our earlier work
\cite{Bishop:2012_honeyJ1-J2} were much more indicative of a direct
first-order transition at $x_{c_{2}}$.  We see no signals at all of
the spiral ordering that is present classically for $x>\frac{1}{6}$
for any value of $x$ in the range $0<x<1$ examined.

\section{Summary}
\label{summary}
\begin{figure}[t]
\begin{center}
\includegraphics[width=8cm]{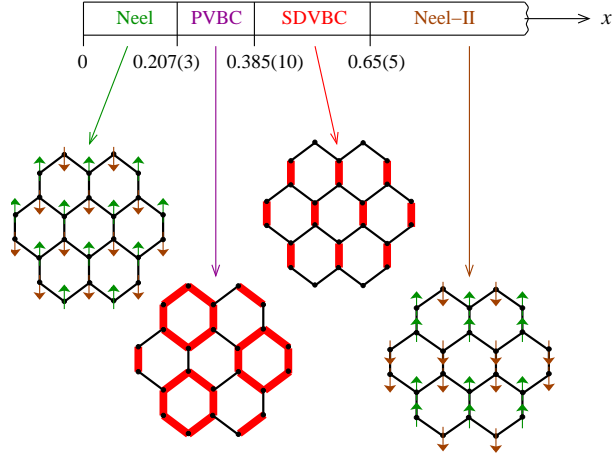}  
\caption{(Colour online) Phase diagram of the spin-$1/2$
  $J_{1}$--$J_{2}$ model on the honeycomb lattice (with $J_{1}>0$ and
  $x \equiv J_{2}/J_{1}>0$), as obtained by a CCM analysis.  The
  quantum critical points are at $x_{c_{1}} \approx 0.207(3)$, $x_{c_{2}} \approx 0.385(10)$, and $x_{c_{3}} \approx 0.65(5)$, as
  shown in the diagram.}
\label{phase}
\end{center}
\end{figure}

In conclusion, over the range $0<x<1$ we find that the spin-1/2
$J_{1}$--$J_{2}$ HAFM on the honeycomb lattice has four phases with,
respectively, N\'{e}el, PVBC, SDVBC, and N\'{e}el-II ordering.  Our CCM estimate for the phase diagram is shown in figure~\ref{phase}.  
We note that all of our most accurate estimates for the three quantum
critical points are based on evaluations of the positions at which the
relevant magnetic order parameters and/or the inverse susceptibilities
to the relevant forms of valence-bond solid order vanish.  Since there
are no energy crossings between the N\'{e}el and N\'{e}el-II states
directly used as CCM model states in our CCM calculations, the GS
energy data only give direct corroborating evidence for the
transitions at $x_{c_{1}}$ and $x_{c_{2}}$ from the corresponding
termination points of the CCM LSUB$m$ solutions based on the N\'{e}el
and N\'{e}el-II model states respectively, as discussed in section
\ref{results} and illustrated in figure \ref{E}.

Our first calculated critical point, $x_{c_{1}} \approx 0.207(3)$, at
which N\'{e}el order melts, agrees well with other recent results,
including $x_{c_{1}} \approx 0.195(25)$ from a large-scale exact
diagonalization (ED) study \cite{Albuquerque:2011_honey}, $x_{c_{1}}
\approx 0.26$ \cite{Zhu:2012_honeyJ1-J2} and $x_{c_{1}} \approx 0.22$
\cite{Ganesh:2013_J1J2honey} from two separate density-matrix
renormalization group (DMRG) studies, and $x_{c_{1}} \approx 0.2075$
\cite{Zhang:2013_J1J2honey} and 0.21 \cite{Yu:2013:honey} from two
recent SB-MFT studies.  Both DMRG studies
\cite{Zhu:2012_honeyJ1-J2,Ganesh:2013_J1J2honey} and the ED study
\cite{Albuquerque:2011_honey} concur with us that the transition at
$x_{c_{1}}$ is probably a continuous deconfined one to a PVBC state,
whereas both SB-MFT studies \cite{Zhang:2013_J1J2honey,Yu:2013:honey}
indicate a transition to a QSL state.

Our second calculated critical point, $x_{c_{2}} \approx 0.385(10)$,
at which the PVBC order melts, is similarly in good agreement with the
result $x_{c_{2}} \approx 0.375(25)$ from the ED study
\cite{Albuquerque:2011_honey}, and the results $x_{c_{2}} \approx
0.36$ \cite{Zhu:2012_honeyJ1-J2} and $x_{c_{2}} \approx 0.35$
\cite{Ganesh:2013_J1J2honey} from the two DMRG studies.  We find that
the transition at $x_{c_{2}}$ is probably a direct first-order one to
a state with SDVBC order.  Both DMRG studies
\cite{Zhu:2012_honeyJ1-J2,Ganesh:2013_J1J2honey} concur that the
transition at $x_{c_{2}}$ is to a state with SDVBC order, although
Ganesh {\it et al}. \cite {Ganesh:2013_J1J2honey} find evidence for
the surprising scenario that the transition at $x_{c_{2}}$ is also of
the continuous deconfined type, as at $x_{c_{1}}$.  The two SB-MFT
studies \cite{Zhang:2013_J1J2honey,Yu:2013:honey} find QSL states out
to values $x \approx 0.3732$ \cite{Zhang:2013_J1J2honey} and $x
\approx 0.43$ \cite{Yu:2013:honey}, respectively, beyond the point
$x_{c_{1}}$ at which N\'{e}el order melts.  They disagree, however,
between themselves as to what is the nature of the GS phase for larger
values of $x$, beyond the QSL phase.  Thus, Zhang and Lamas
\cite{Zhang:2013_J1J2honey} find the GS phase to be spirally ordered
(just as in the classical, $s \rightarrow \infty$, version of the
model) for $0.398 \lesssim x\; (\lesssim 0.5)$, and to have SDVBC order in
the very narrow region $0.3732 \lesssim x \lesssim 0.398$; whereas Yu
{\it et al}. \cite{Yu:2013:honey} find that for $x \gtrsim 0.43$ the GS
phase has N\'{e}el-II order.  The ED study
\cite{Albuquerque:2011_honey}, by contrast, finds a first-order
transition at $x_{c_{2}}$ to a state that cannot be distinguished
between having either SDVBC or N\'{e}el-II order.

Finally, we find evidence for a third critical point at $x_{c_{3}}
\approx 0.65(5)$ at which a continuous (and hence again deconfined)
transition occurs to a state with weak N\'{e}el-II magnetic order.  We
note that such a transition is also compatible with the DMRG result of
Ganesh {\it el al}. \cite{Ganesh:2013_J1J2honey}, which could not rule
out a melting of the SDVBC order for values $x \gtrsim 0.7$.  It is
interesting to speculate whether the weak N\'{e}el-II magnetic order
observed by us for $x > x_{c_{3}}$ might be interpreted as, or arise
from, a sort of ``dressed'' SDVBC state in which spin-triplets now
contribute on the spin-singlet dimer bonds.  It is too far beyond the
scope of the present analysis, however, to address such delicate
questions authoritatively.

As a last remark, it is interesting to note that in a very recent
study using a projector QMC technique \cite{Damle:2013_honey} a very
similar direct continuous quantum phase transition to what we observe
here for the $J_{1}$--$J_{2}$ model at $x_{c_{1}}$, between states
with N\'{e}el and PVBC order, has also been observed in a related
spin-1/2 $J_{1}$-$Q$ model on the honeycomb lattice, of precisely the
type suggested by Yang {\it et al}. \cite{Yang:2012} to be more
relevant to the low-energy physics of the FH model on the honeycomb
lattice, as discussed previously in section~\ref{intro}.  This
$J_{1}$-$Q$ model also contains NN AFM exchange bonds of strength
$J_{1}$, but with our competing NNN exchange bonds of strength $J_{2}$
replaced by a six-spin interaction term of strength $Q$ on hexagonal
plaquettes, which by itself favours the formation of a state with PVBC
order.  It would clearly also be of interest to apply a comparable CCM
study to the $J_{1}$-$Q$ model to that used here for the
$J_{1}$--$J_{2}$ model, in order to investigate its GS phase diagram
similarly.

\section*{ACKNOWLEDGMENT}
We thank the University of Minnesota Supercomputing Institute for the
grant of supercomputing facilities for this research.

\end{document}